\documentclass[11pt]{article}

\usepackage{latexsym}
\usepackage[dvips]{graphicx}

\usepackage{amsmath,amssymb}

\newtheorem{thm}{Theorem}
\newtheorem{df}{Definitions}
\newtheorem{lm}{Lemma}

\newtheorem{pr}{Proposition}

\newcommand{\qed}{\hspace*{\fill} $\Box$}

\title{How to solve the cake-cutting problem in sublinear time --- ver.~2}

\author{
Hiro Ito\thanks{
     School of Informatics and Engineering, 
    The University of Electro-Communications (UEC), 
    Tokyo, Japan; and
    CREST, JST, Tokyo, Japan; 
    {\tt itohiro@uec.ac.jp}
    }
\and 
Takahiro Ueda\thanks{
       Komatsu Ltd., Tokyo, Japan; 
    {\tt mx.u.2147483647@gmail.com}
    }
 }

\begin{document}

\setcounter{page}{0}

\maketitle

\begin{abstract}
The cake-cutting problem refers to the issue of dividing a cake into pieces and distributing them to players who have different value measures related to the cake, and who feel that their portions should be ``fair.'' The fairness criterion specifies that in situations where $n$ is the number of players, each player should receive his/her portion with at least $1/n$ of the cake value in his/her measure. In this paper, we show algorithms for solving the cake-cutting problem in sublinear-time. More specifically, we preassign fair portions to $o(n)$ players in $o(n)$-time, and minimize the damage to the rest of the players. All currently known algorithms require $\Omega(n)$-time, even when assigning a portion to just one player, and it is nontrivial to revise these algorithms to run 
in $o(n)$-time since many of the remaining players, who have not been asked any queries, may not be satisfied with the remaining cake. To challenge this problem, we begin by providing a framework for solving the cake-cutting problem in sublinear-time. Generally speaking, solving a problem in sublinear-time requires the use of approximations. However, in our framework, we introduce the concept of ``$\epsilon n$-victims,'' which means that $\epsilon n$ players (victims) may not get fair portions,  where $0< \epsilon \leq 1$ is an arbitrary constant. In our framework, an algorithm consists of the following two parts: In the first (Preassigning) part, it distributes fair portions to $r < n$ players in $o(n)$-time. In the second (Completion) part, it distributes fair portions to the remaining $n-r$ players except for the $\epsilon n$ victims in $\mbox{poly}(n)$-time. There are two variations on the $r$ players in the first part. Specifically, whether they can or cannot be designated. We will then present algorithms in this framework. In particular, an $O(r/\epsilon)$-time algorithm for $r \leq \epsilon n/127$ undesignated players with $\epsilon n$-victims, and an $\tilde{O}(r^2/\epsilon)$-time algorithm for $r \leq \epsilon e^{{\sqrt{\ln{n}}}/{7}}$ 
designated players and $\epsilon \leq 1/e$ with $\epsilon n$-victims are presented.
\end{abstract}

\newpage
\section{Introduction}

\subsection{What is a sublinear-time algorithm for the cake-cutting problem?}

This paper reports the first results 
on sublinear-time algorithms for solving the cake-cutting problem, in which it is necessary to divide a given cake into pieces and to distribute those pieces to players in a way that ensures that all players are ``satisfied,'' more specifically, in a way that ensures every player believes that his/her portion has at least $1/n$ value of the whole cake, where $n$ is the number of the players. It has previously been known that an $O(n \log n)$-time algorithm~\cite{art2} and that a $\Theta (n \log n)$-time lower-bound exists for deterministic algorithms~\cite{art5}. For approximation, a linear-time $O(1)$ approximation algorithm has been given~\cite{art5}. 

However, when we divide a property and assign portions to persons, sometimes situations arise in which it is impossible to meet the requirements of all interested persons simultaneously. In such cases, it may be necessary to assign acceptable portions to some persons without asking for approval from all other interested persons. This is the motivation behind our desire to develop algorithms for solving the cake-cutting problem in sublinear time. 

Herein, we consider ways to {\em preassign portions to a number of ($r=o(n)$) players in $o(n)$-time}. However, since all the known algorithms need $\Omega (n)$-time, even for assigning a portion to just one player, the problem is nontrivial. In fact, it is a difficult matter to satisfy just one player. Moreover, even if $r$ players can be satisfied, if the other $n-r$ players are unsatisfied, the solution is clearly suboptimal in many cases. Thus, it is better to be able to satisfy the remaining players by distributing the remaining cake appropriately. However, it is often very hard (or even impossible) to completely satisfy all the other players since we have already assigned a portion of the cake 
after giving queries to only a sublinear number of players.  Thus, we need to make some approximations. With this in mind, we hereby introduce the concept of ``$\epsilon n$-victims,'' which means that {\em we can give up trying to satisfy at most $\epsilon n$ players (victims)}.

Recently, it has been learned that many problems can be approximated in sublinear time \cite{BSS_MC-testable_STOC08,PropertyTestingLNCS10,GR-STOC97,GGR-JACM98,HKNO_LocalPartition_FOCS09,ItoYoshida_Knapsack_TAMC12,ItoTanigawaYoshida_ICALP12,NS_Testable_STOC11,Levi-Ron_PO_ICALP13,Yoshida_STOC14}. To solve a problem in sublinear time, it is necessary to introduce some approximations by using a parameter $0< \epsilon  \leq 1$.  There are two types of approximations. In the first, which is for decision problems, an edit distance between an instance and an objective property is defined and algorithms distinguish between instances satisfying the property and those $\epsilon$-far from the property with high provability. The second type, which is for optimizing problems, provides $\epsilon n$-approximation solutions for objective functions~\cite{PropertyTestingLNCS10}.  

All known sublinear-time algorithms use one of these two approximation types. Our approximation, $\epsilon n$-victims, can be seen as the latter approximation type. Thus, if we stipulate that the goal when solving the cake-cutting problem is to maximize the number of satisfied players, a solution with $\epsilon n$-victims is an $\epsilon n$-approximation solution.

The results of this paper can be summarized as follows:
\begin{itemize}
\item Presenting a framework for solving the cake-cutting problem in sublinear-time.
\item Presenting sublinear-time algorithms under this framework.
\end{itemize}

The framework presented here is as follows: 
\\
\noindent
\textbf{The proposed sublinear-time cake-cutting framework:} 
\begin{itemize}
\item[(1)] \textbf{First (Preassigning) Part:} First, we preassign portions to $r=o(n)$ players in $o(n)$-time.
\item[(2)] \textbf{Second (Completion) Part:} 
Next, we assign portions to the remaining $n-r$ players except for the $\epsilon n$ victims in $\mbox{poly}(n)$-time.
\qed
\end{itemize}

Note that it is impossible to do the second part in sublinear-time, since it is necessary to assign one portion to each of the remaining $\Omega (n)$ players (except for the $\epsilon n$ victims). 

Next, we will consider sublinear-time cake cutting algorithms that obey this framework. These algorithms can be divided into two types: one in which the \textbf{the preassigned players cannot be designated}, and the other in which textbf{they can be designated}. We will then present algorithms for both types. More specifically, for the first (undesignated) type, we can preassign portions to the $r \leq {\epsilon n}/{127}$ undesignated players in $O({tr}/{\epsilon})$-time and set the success probability to at least $1 - (\frac{1}{64})^{t/\epsilon} - \frac{8}{(2t-3)^2 r}$. After that, we can assign portions to the remaining players except for the $\epsilon n$ victims in $O(n \log{n})$-time, where $t \geq 1$ is an arbitrary real number.
%
For the latter (designated) type,
for any $0< \epsilon \leq 1/e$, we can preassign portions to $r \leq \epsilon e^{{\sqrt{\ln{n}}}/{7}}$ designated players in $\tilde{O}({tr^2}/{\epsilon})$-time and set the success provability to at least $1- (\epsilon/r)^t \geq 1-e^{-t}$. After that, we can then assign portions to the remaining players except for the $\epsilon n$ victims in $O(rn \log{rn})$-time, where $t \geq 1$ is an arbitrary real number.

\subsection{Definition of the cake-cutting problem}
Let $P$ be the set of $n$ players. We assume that every algorithm for solving the cake-cutting problem knows $n$ (which is the number of players)\footnote{Although this assumption may seem trivial, it is an important consideration in sublinear-time algorithms since the algorithm cannot count the number of players in sublinear-time. Such an assumption is generally introduced when sublinear-time algorithms are investigated.}. 

The cake is represented by the unit interval $C=[0,1]$. The portion of each player is a set of disjointed subintervals of $C$. Every player $p \in P$ has his/her subjective nonnegative value function $\mu_p: 2^C \to [0,1]$, which is defined on every measurable subset of $C$. Furthermore, $\mu_p$ is additive. In other words, the value of the portion of a player is the sum of the subinterval values of his/her portion. The value function is normalized, i.e., $\mu_p(C) = 1$ for every $p \in P$.
%

A portion $C_p \subseteq C$ of a player $p \in P$ is deemed to be {\em fair} if $\mu_p (C_p) \geq 1/n$. For any positive real $c \geq 1$, $C_p$ is deemed to be {\em $c$-fair} if $\mu_p (C_p) \geq 1/cn$.

When evaluating cake-cutting algorithms, the Robertson-Webb model~\cite{book1} is generally used. In this model, the two query types listed below are allowed, and the complexity of an algorithm is evaluated by the {\em query complexity}. In other words, the number of these queries made by the algorithm.

\begin{itemize}
\item {\em Cut query}: 
For a continuous piece of a cake $D=[a,b] \subseteq C$ ($0 \leq a < b \leq 1$), a player $p \in P$, and a positive real number $0 \leq \alpha \leq 1$, a query $\mbox{\sc Cut}(D, p, \alpha)$ returns the smallest value $x \in [a,b]$ such that $\mu_p([a,x]) = \alpha$. If there is no such $x$ (i.e., $\mu_p(D) < \alpha$), then it returns some predefined massage.

\item {\em Evaluation query}: 
For a continuous piece of a cake $D=[a,b] \subseteq C$ ($0 \leq a < b \leq 1$), a player $p \in P$, and a positive real number $x \in [a,b]$ ($=$ a point on $D$), a query {\sc Eval}$(D, p, x)$ returns the value $0 \leq \alpha \leq 1$ such that $\mu_p([a,x]) = \alpha$. If $x \geq b$ (i.e., asking to evaluate whole of $D$), then {\sc Eval}$(D, p, x)$ is simply expressed as {\sc Eval}$(D, p)$. 
\end{itemize}

\subsection{Previous work}
One well known method of handing the cake-cutting problem, involves an $O(n \log{n})$-time algorithm using the divide-and-conquer concept~\cite{art2}. This algorithm divides a cake into two pieces and assigns players into two half-sized subsets. It begins by assigning one of the pieces to one of the subgroups, and the other piece to the other subgroup. Then, it recursively applies this separation until every subgroup becomes a singleton. Note that this algorithm requires $\Theta(n)$ queries even if it only assigns a portion to one player. We refer to this algorithm as {\sc DC}$(P, C)$.

For the lower-bound results, it is known that the time-complexity is $\Theta(n \log n)$ for deterministic algorithm~\cite{art5}. In the same paper, they showed that, with some restrictions, this bound can be also applied to a randomized case. For approximations, Edmonds and Pruhs~\cite{art6} showed that for $c$-fair division with $c > 32$, there is an $O(n)$-time randomized algorithm. 

The success probability of this algorithm is at least
\begin{equation}
1-\frac{2^{13}}{c^2 (c - 32)} - \frac{1024}{c^3} - \frac{128}{c^2} \label{eq:AF}
\end{equation} for $c > 32$.  We refer this algorithm as {\sc ApproxFair}$(P, C, c)$.
Our algorithms use these algorithms as subroutines. 
In addition to these two, numerous other algorithms have been presented~\cite{art11,art1}.  
If it is necessary to distribute pieces to $\Omega (n)$ players fairly, clearly we need $\Omega (n)$-time. 
However, none of the previously known algorithms have considered preassigning portions to $o(n)$ players, and they all need $\Omega(n)$ queries even when assigning a portion to just one player.

\subsection{Our results}

In this subsection, we will explain the results we have obtained thus far. Throughout this paper, we assume that every player is honest, i.e., that he/she gives correct answers for every query\footnote{Even if there is a dishonest player, the honest players will get fair portions, while the dishonest player may not.}. 
First we show a preliminary result, which can be obtained as a simple application of \cite{art5}, as follows: 
\begin{pr}\label{pr:basic}
For any $t \geq 64$ and any given subset $P_r \subseteq P$ of players with $|P_r| = r \leq {n}/{t}$, there is an $O(r)$-time algorithm for assigning fair portions to all players in $P_r$ 
with success probability at least $1 - 2^9/t^2$.
\end{pr}
The complexity of this algorithm is $O(r)$, and it is clearly the best possible because it matches with the trivial lower bound. Moreover, it also allows us to arbitrarily assign designated $r$ players. However, one obvious flaw of this algorithm is that it may victimize all of the remaining players, so efforts should be made to reduce the number of victims. The following algorithm allows a maximum of $\epsilon n$ victims for any given $0 < \epsilon \leq 1$:
\begin{thm}\label{th:U1}
For any positive real number $0 < \epsilon \leq 1$, any positive integer $r \leq {\epsilon n}/{127}$, and any constant real number $t > {3}/{2}$, there is an algorithm for preassigning fair portions to $r$ players in $O({tr}/{\epsilon})$-time, and then assigning fair portions to the remaining players except for the $\epsilon n$ players (victims) in $O(n \log{n})$-time  with success probability at least $1 - \frac{8}{(2t-3)^2r} - (\frac{1}{64})^{{t}/{\epsilon}}$.
\end{thm}

While in the algorithm of Theorem~\ref{th:U1}, 
preassigned $r$ members cannot be designated, the following shows an algorithm in which they can:
\begin{thm}\label{th:S1}
For any real numbers $0 < \epsilon \leq 1/e$ and $t \geq 1$, and any set of $r \leq \epsilon e^{{\sqrt{\ln{n}}}/{7}}$ players $P_r$, there is an algorithm for preassigning fair portions to all players in $P_r$ in $O(\frac{tr^2}{\epsilon} (\log{\frac{r}{\epsilon}})^3)$-time and then assigning fair portions to remaining players except for the $\epsilon n$ players (victims) in $O(r n \log(r n))$-time 
with success probability at least $1- (\epsilon/r)^t \geq 1 - e^{-t}$.
\end{thm}

\subsection{Organization}

The remainder of this paper is organized as follows:
In Section~\ref{sc:basic}, we present a proof of Proposition~\ref{pr:basic}. In Sections~\ref{sc:undes} and \ref{sc:des}, we examine the undesignated version (Theorem~\ref{th:U1}) and the designated version (Theorem~\ref{th:S1}), respectively. In Section~\ref{sc:summary}, we summarize our results and discuss on future work.

\section{Proof of Proposition~\ref{pr:basic}}\label{sc:basic}

In this short section, we show the following proof of Proposition~\ref{pr:basic}.\\

\noindent
{\em Proof of Proposition~\ref{pr:basic}}: 
It is sufficient to simply call $\mbox{\sc ApproxFair}(r, P_r, C, t)$. It assigns $t$-fair portions to all players in $P_r$. In other words, they fell at least $\frac{1}{tr}$ value in their own portion. 
From the assumption $r \leq \frac{n}{t}$ it follows that $\frac{1}{tr} \geq \frac{1}{n}$. 
Therefore, they all get fair portions, and by considering (\ref{eq:AF}) and $t \geq 64$, the probability of failing is at most
\begin{eqnarray*}
\frac{2^{13}}{t^2 (t - 32)} + \frac{2^{10}}{t^3} + \frac{2^7}{t^2}
\leq \frac{2^{14}}{t^3} + \frac{2^{10}}{t^3} + \frac{2^7}{t^2}
\leq \frac{2^{8}}{t^2} + \frac{2^{4}}{t^2} + \frac{2^7}{t^2} 
\leq \frac{2^9}{t^2}.
\end{eqnarray*}
The time-complexity is clearly $O(r)$. 
\qed

\section{Undesignated $r$ players}\label{sc:undes}

In this section, we consider a case where $P_r$ cannot be designated. 

\subsection{Algorithm for Theorem~\ref{th:U1}}

When preassigned players are not designated, the algorithms can select $P_r$ players arbitrarily. That is, players who feel a relatively high value in a specified part (e.g., the left-side part of the cake) are considered more suitable, such members can be selected at high probability levels by using asking cut-queries to a number ($\lceil {tr}/{\epsilon} \rceil$) of other players. Let $P'$ be the set of selected players and let $C'$ be a piece to which these players in $P'$ have assigned high value ($128r/n$). Then, by applying {\sc ApproxFair} to $P'$ and $C'$ with approximation parameter $128$, the players in $P'$ have a high probability of getting fair portions. This summarizes the preassigning part. 

For the completion part, it can be expected that a small number of the remaining players will feel that $C'$ (the removed piece) has high value, and that the only way the remaining players can share the rest of the cake ($C-C'$) fairly is by removing the appropriate $\epsilon n$ players (victims).

Before showing the details of this algorithm, we will first define a subroutine {\sc Pcut} it uses. The objective of this subroutine is to get a set of $m \in \{ 0, \ldots, n \}$ players from $Q \subseteq P$ who have a high probability of seeing relatively high value in the left-most part of the piece $D$. \\

\noindent
\textbf{procedure} {\sc Pcut}($Q,D,\alpha, m$)\\
\textbf{Input:} $Q \subseteq P$, $D \subseteq C$, real value $0 \leq \alpha \leq 1$, integer $0 \leq m \leq n$; \\
\textbf{begin}\\
1 \hspace{1em} \textbf{for} $p \in Q$ \textbf{do}\\
2 \hspace{2em} $x_p := \mbox{\sc Cut}(D,p,\alpha)$\\
3 \hspace{1em} \textbf{enddo}\\
4 \hspace{1em} Let $Q'$ be the set of players $p \in Q$ having the 1st, 2nd, $\ldots$, and the $m$th smallest value $x_p$ in $Q$, \\ \hspace{2em} where ties are broken arbitrarily.\\
5 \hspace{1em} \textbf{Output} $Q'$ \\
\textbf{end.}\\

The preassigning part of the algorithm used for proving Theorem~\ref{th:U1} is as follows: \\

\noindent
\textbf{procedure} {\sc PreassignU}($P,C,r, \epsilon, t$)\\
\textbf{Input:} The set $P$ of $n$ players, The cake $C=[0,1]$, positive integers $r$ and $t$,  real value $0 < \epsilon \leq 1$;\\ 
\textbf{begin}\\
01 \hspace{1em} $P_0 := \emptyset$\\
02 \hspace{1em} \textbf{for} $\lceil \frac{tr}{\epsilon} \rceil$ times \textbf{do}\\
03 \hspace{2em} Select $p \in P$ UAR and $P_0 := P_0 \cup \{ p \}$;\\
04 \hspace{1em} \textbf{enddo}\\
05 \hspace{1em} \textbf{if} $|P_0| < r$ \textbf{then}  \textbf{output} ``Failed'' and \textbf{stop}  \textbf{endif};\\
06 \hspace{1em} $P' := \mbox{\sc Pcut}(P_0, C, \frac{128r}{n}, r)$\\
07 \hspace{1em} $x := \max_{p \in P'} x_p$\\
08 \hspace{1em} $C':= [0,x]$\\
09 \hspace{1em} \textbf{for} $\lceil \frac{t}{\epsilon} \rceil$ times \textbf{do}\\
10 \hspace{2em} \textbf{call} {\sc ApproxFair}($r,P',C',128$)\\
11 \hspace{2em} \textbf{if} above {\sc ApproxFair} succeeds \textbf{then}\\
12 \hspace{3em} \textbf{output} the assignment obtained in Line~10; \textbf{stop};\\
13 \hspace{2em} \textbf{endif};\\
14 \hspace{1em} \textbf{enddo}\\
15 \hspace{1em} \textbf{comment} all {\sc ApproxFair} in Line 10 failed;\\
16 \hspace{1em} \textbf{output} ``Failed'';\\
\textbf{end.}\\

By applying {\sc PreassignU}, providing it does not fail, all players in $P'$ ($|P'|=r$) will have their own portions (which we will later prove are fair). The next important point is ensuring that the remaining players except for $\epsilon n$ victims will be satisfied. We define the other terms used for treating this problem as follows: 

\begin{df}
Let $Q \subseteq P$ and $D \subseteq C$ be a subset of players and a subset of the cake, respectively. 
A player $p \in Q$ is called {\em safe with respect to $(Q,D)$} if $\mu_p(D) \geq \frac{|Q|}{n}$, or {\em dangerous with respect to $(Q,D)$} otherwise. 
We may omit ``with respect to $(Q,D)$'' if $(Q,D)$ is clear. 
If all players in $Q$ are safe with respect to $(Q,D)$, we then say that $Q$ is {\em safe with respect to $D$}, or {\em safe} in short, if $D$ is clear. 
For $m \geq 0$, if there is a subset of $Q' \subset Q$ such that $|Q'| \leq m$ and $Q-Q'$ is safe, then $Q$ is called {\em $m$-safe}.  
\end{df}

If $Q$ is safe with respect to $D$, it is clear that all players in $Q$ can get fair portions in $D$ 
by using arbitrary cake-cutting-algorithms, such as {\sc DC}$(Q,D)$ (Lemma~\ref{lm:DC}, which will be shown later). Then, for proving the completion part following {\sc PreassignU}, we should show that $P-P'$ is $\epsilon n$-safe with respect to $C-C'$. The algorithm of the completion part is simple: It is sufficient to make a query {\sc Eval}($C-C',p$) for every player $p$ in $P-P'$ and remove the lowest evaluating $\epsilon n$ players. 
Pseudo code 
of this algorithm is shown below:\\ 

\noindent
\textbf{procedure} {\sc Completion}($P-P',C-C',n, \epsilon$)\\
\textbf{begin}\\
01 \hspace{1em} $Q := \mbox{\sc Victimize}(P-P',C-C', \lfloor \epsilon n \rfloor )$\\
02 \hspace{1em} \textbf{call} {\sc DC}($Q,C-C'$)\\
\textbf{end.}\\

\noindent
\textbf{procedure} {\sc Victimize}($P'',D,m$)\\
\textbf{Input:} Subset $P'' \subseteq P$ of players, subset $D \subseteq C$ of the cake, integer $m \geq 0$;\\
\textbf{begin}\\
1 \hspace{1em} \textbf{for} $p \in P''$ \textbf{do} $x_p := \mbox{\sc Eval}(D,p)$ 
\textbf{enddo} \\
2 \hspace{1em} Let $Q_{\mbox{\scriptsize vict}} \subseteq P''$ be the set of $m$ players having the 1st, 2nd, $\ldots$, $m$th smallest values of $x_p$, \\
\hspace{2em} where ties are broken arbitrarily;\\
3 \hspace{1em} \textbf{output} $Q := P'' - Q_{\mbox{\scriptsize vict}}$; \\
\textbf{end}.\\


\subsection{Proof of Theorem~\ref{th:U1}}

We prepare the following lemmas for showing the proof of Theorem~\ref{th:U1}. 

\begin{lm}\label{lm:U1-2}
Let $N$ be $\{1,2,\ldots,n\}$ and $S$ be an $\lfloor \epsilon n \rfloor$ size subset of $N$ for $0 < \epsilon \leq 1$. For real numbers $s, t > 1$ such that $(s - 1) (t - 1) > 1$ and a positive integer $r$ such that $r \leq {\epsilon n}/{s}$, if we choose at least ${tr}/{\epsilon}$ elements from $N$ uniformly at random (UAR), we then get at least $r$ different elements in $S$ with probability at least $1-\frac{s^2}{((s - 1) (t - 1) - 1)^2 r}$.
\end{lm}

\noindent
{\em Proof}: 
Let $X$ be the random variable of the number of chosen elements until we get $r$ different elements in $S$ from $N$. Further, let $X_i$ be the random variable of the number of chosen elements until we get $i$-th different elements in $S$ after $i-1$ different elements were chosen from $S$. Clearly, $X= \sum_{i=1}^{r}X_i$.

Let $p_i$ be the probability that we get a new element from $S$ after we have gotten $i - 1$ different elements from $S$. The following inequalities hold:
\[
p_i = \frac{\lfloor \epsilon n \rfloor -( i - 1)}{n} 
\geq \frac{\lfloor \epsilon n \rfloor - (r - 1)}{n} 
> \frac{\epsilon n - r}{n} 
\geq \frac{\epsilon n - \frac{\epsilon n}{s}}{n}\ 
= \frac{(s - 1)\epsilon}{s}.
\]
Since the random variable $X_i$ follows a geometric distribution, the expected value $E[X_i]$ and the variance $V[X_i]$ satisfy $E[X_i] = {1}/{p_i}$ and $V[X_i] = {(1-p_i)}/{p_i^2}$, respectively. By the linearity of expected value, 
\[
E[X] = \sum_{i=1}^rE[X_i] \leq \sum_{i=1}^r\frac{s}{(s-1)\epsilon} = \frac{sr}{(s - 1)\epsilon}.\]

Since each $X_i$ is independent, the variance satisfies linearity, and thus
\[
V[X] = \sum_{i=1}^rV[X_i] 
\leq \sum_{I = 1}^r \frac{1 - \frac{(s - 1) \epsilon}{s}} {\left(\frac{(s - 1)\epsilon}{s} \right)^2} 
= \frac{(s - \epsilon s+\epsilon)sr}{(s - 1)^2\epsilon^2}.
\]
We compute the probability that we do not get at least $r$ different elements in $S$ when we choose ${tr}/{\epsilon}$ elements from $N$ uniformly, at random, as follows:
 
\begin{eqnarray*}
\mbox{\rm Pr}\left[X>\frac{tr}{\epsilon}\right] 
&\leq& \mbox{\rm Pr}\left[X\geq\frac{tr}{\epsilon}\right] 
\leq \mbox{\rm Pr}\left[ \left|X-\frac{sr}{(s-1)\epsilon} \right| \geq \frac{tr}{\epsilon}-\frac{sr}{(s-1)\epsilon}\right]\\
 &=& \mbox{\rm Pr}\left[ \left| X-\frac{sr}{(s-1)\epsilon} \right| 
 \geq \frac{sr}{(s-1)\epsilon} \left( t\frac{s-1}{s}-1 \right) \right]\\
 &\leq& \mbox{\rm Pr}\left[|X-E[X]| \geq \frac{sr}{(s-1)\epsilon} \left( t\frac{s-1}{s}-1 \right) \right].\\
&&\mbox{(From Chebyshev bound, 
$\mbox{\rm Pr}[| X - E[X] | \geq a] \leq \frac{V[X]}{a^2},~\forall a>0$)}\\
 &\leq& \frac{V[X]}{\left(\frac{sr}{(s-1)\epsilon}(t\frac{s-1}{s}-1)\right)^2}\\
 &\leq& \frac{\frac{(s - \epsilon s + \epsilon) s r}{(s - 1)^2 \epsilon^2}}
 {\left(\frac{s r}{(s - 1) \epsilon}(t \frac{s - 1}{s} - 1)\right)^2} = \frac{s-(s-1)\epsilon}{(t\frac{s-1}{s}-1)^2 rs}\\
 &\leq& \frac{s}{\left(t\frac{s-1}{s}-1\right)^2 rs} = \frac{s^2}{(st-s-t)^2 r} = \frac{s^2}{((s-1)(t-1)-1)^2 r}.
\end{eqnarray*}
The desired inequality is obtained. 
\qed

\begin{lm}\label{lm:U1-3}\label{lm:DC}
For any $Q \subseteq P$ and $D \subseteq C$, if $Q$ is safe with respect to $D$, then all players in $Q$ can be get fair portions in $D$ by using arbitrary cake-cutting algorithms.
\end{lm}

\noindent
{\em Proof}: 
By applying a cake-cutting algorithm, every player $p \in Q$ obtains a portion with value at least ${\mu_p(D)}/{|Q|}$. From that, $Q$ is safe with respect to $D$, $\mu_p(D) \geq |Q|/n$ for $\forall p \in Q$. Thus, the value of the cake obtained by $\forall p \in Q$ is 
$$
\frac{\mu_p(D)}{|Q|} \geq \frac{1}{|Q|} \cdot \frac{|Q|}{n} = \frac{1}{n}. 
$$
\qed\\

\noindent
{\em Proof of Theorem~\ref{th:U1}}: 
Next, we will show the following facts: 
\begin{itemize}
\item[(i)] All players in $P'$ get fair portions with probability 
at least $1 - \frac{8}{(2t-3)^2 r} - \left(\frac{1}{64} \right)^{{t}/{\epsilon}}$ after calling {\sc ApproxFair} in line 10 of {\sc PreassignU}. 
\item[(ii)]  $Q$ (the output of $\mbox{\sc Victimize}(P-P',C-C',  \lfloor \epsilon n \rfloor )$ 
in line 01 of {\sc Completion}($P-P', C-C', n, \epsilon $) is safe with respect to $C-C'$. 
\end{itemize}

In what follows, we show proofs for the above items. 
\begin{itemize}
\item[(i)] 
First, we assume that $|P_0| \geq r$ in line 05 of {\sc PreassignU} and that at least one call of {\sc ApproxFair} in line 10 of {\sc PreassignU} succeeds. Let $C_p$ be the portion that player $p \in P'$ gets by this {\sc ApproxFair} when it succeeds.  
From the property of {\sc ApproxFair}, $C_p$ is at least 128-fair, i.e., $\mu_p(C_p) \geq {\mu_p(C')}/{128r}$. 
From the operations in lines 06--08 of {\sc PreassignU}, $\mu_p(C') \geq {128r}/{n}$. 
It then follows that $$\mu_p(C_p) \geq  \frac{1}{128r} \cdot \frac{128r}{n} = \frac{1}{n},$$ i.e., each player in $P'$ gets a fair portion.  

Next, we estimate the probability that $|P_0| \geq r$ in line 05 of {\sc PreassignU} and that at least one call of {\sc ApproxFair} in line 10 of {\sc PreassignU} succeeds. From Lemma~\ref{lm:U1-2} with regarding $P$ and $\mbox{\sc Pcut}(P,C, \frac{128r}{n}, \lfloor \epsilon n \rfloor)$ as $N$ and $S$, respectively \footnote{The reason that $\mbox{\sc Pcut}(P,C, \frac{128r}{n}, \lfloor \epsilon n \rfloor)$ is considered here is explained in (ii).} and by letting $s=127$, it follows that the probability that $|P_0 \cap \mbox{\sc Pcut}(P,C, \frac{128r}{n}, \lfloor \epsilon n \rfloor)| < r$ 
occurs is at most $\frac{127^2}{(126t-127)^2 r}$.  
From the assumption of $t > 3/2$, this probability becomes 
$$
\frac{127^2}{(126t-127)^2 r} < \frac{(127/126)^2}{(t-3/2)^2 r} < \frac{8}{(2t-3)^2 r}. 
$$

$|P_0 \cap \mbox{\sc Pcut}(P,C, \frac{128r}{n}, \lfloor \epsilon n \rfloor)| \geq r$ includes $|P_0| \geq r$ and $|P'| =r$. 

From (\ref{eq:AF}), the probability that one call of {\sc ApproxFair} in line 10 of {\sc PreassignU} succeeds is at least $$1  - \frac{2^{13}}{2^{14} (128-32)} - \frac{1024}{128^3} - \frac{128}{128^2} = 1- \frac{83}{6144} > 1- \frac{1}{64},$$. Thus, the probability that all the calls of {\sc ApproxFair} fail is 
at most $64^{-t/\epsilon}$. 

Therefore, the success probability of this algorithm is at least $$1- \frac{8}{(2t-3)^2 r} - \left( \frac{1}{64} \right)^{{t}/{\epsilon}}.$$

\item[(ii)] 
Assume that $|P_0 \cap \mbox{\sc Pcut}(P,C, \frac{128r}{n}, \lfloor \epsilon n \rfloor)| \geq r$. From this,  $P' \subseteq  \mbox{\sc Pcut}(P,C, \frac{128r}{n}, \lfloor \epsilon n \rfloor)$ follows. This means that for every player $p \in P- \mbox{\sc Pcut}(P,C, \frac{128r}{n}, \lfloor \epsilon n \rfloor)$, $\mu_p(C-C') \geq \frac{|P|-128r}{|P|}$. From the assumption of $r \leq \lfloor {\epsilon n}/{127} \rfloor$ ($\because$ $r$ is an integer), $$ \mu_p(C-C') \geq \frac{|P|-128r}{|P|} > \frac{|P|- \lfloor \epsilon n \rfloor - r}{|P|}. 
$$
It follows that  $Q \subseteq P-\mbox{\sc Pcut}(P,C, \frac{128r}{n}, \epsilon n)$ and $|Q| = n - \epsilon n -r$. Therefore, $Q$ is safe with respect to $C-C'$.  
\end{itemize}

From (ii) and Lemma~\ref{lm:U1-3}, {\sc DC} in line 02 of {\sc Completion} assigns fair portions to all players in $P-P'$. The query complexity of {\sc PreassignU} is clearly $O({tr}/{\epsilon})$. The query complexity of {\sc Completion} is $O(n \log n)$, since 
{\sc DC} can be done in $O(n \log n)$.
\qed

\section{Designated $r$ players}\label{sc:des}

\subsection{Algorithm for Theorem~\ref{th:S1}}

In this section, we consider the case where $P_r$ is given. The key to solving this problem is to find a piece $C_p$ that a player $p \in P_r$ prefers. After finding $C_p$ for all $p \in P_r$, if all $C_p$ are disjointed, we then assign $C_p$ to $p$. Otherwise, i.e., when some $C_{p_1}$, $\ldots$, $C_{p_k}$ are ``connected'' (the definition is given later), we allot $C_{p_1} \cup \ldots \cup C_{p_k}$ to $\{ p_1, \ldots, p_k \}$ by using a suitable cake-cutting algorithm, e.g., {\sc DC}. 

The basic strategy used to find $C_p$ is as follows. In the beginning, $C_p := C$ (of course, it will be trimmed). We ask a randomly chosen constant number of players (let $P_p$ be the set of chosen players) to evaluate $C_p$. If a small number of players evaluate it as high, then $C_p$ is fixed. Otherwise (in the first iteration, this case must occur since $C_p=C$), we divide $C_p$ into two pieces such that the half of players in $P_p$ prefer one of the half pieces and the other players prefer the other piece, and let $C_p$ be the half piece that $p$ prefers. By iteratively applying the above operations some fixed number of times, we have a high probability of getting an appropriate $C_p$. 

To show the details of the first (preassigning) part, we use the following concept. Let ${\cal C} = \{ C_1, \ldots, C_{|{\cal C}|} \}$ be a family of cake subsets. We define the {\em relation graph} $G_{\cal C} = ({\cal C}, E_{\cal C} )$ with respect to ${\cal C}$ as $(C_i,C_j) \in E_{\cal C}$ iff $C_i \cap C_j \neq \emptyset$ for $i,j \in \{ 1, \ldots, |{\cal C}| \}$ and $i \neq j$. \\

\noindent
\textbf{procedure} {\sc PreassignS}($P,C,P_r, \epsilon, t$)\\
\textbf{Input:} The set $P$ of $n$ players, 
The cake $C=[0,1]$, a subset of $r$ players $P_r \subseteq P$, positive integer $t$, real value $0 < \epsilon \leq 1$;\\
\textbf{begin}\\
01 \hspace{1em} \textbf{for all} $p \in P_r$ \textbf{do}\\
02 \hspace{2em} $C_p := \mbox{\sc Deposit}(p,P,C,\epsilon/r,t)$\\
03 \hspace{1em} \textbf{enddo} \\
04 \hspace{1em} Construct the relation graph $G_{\cal C}$ with respect to ${\cal C} := \{ C_p ~|~ p \in P_r \}$.\\
05 \hspace{1em} \textbf{for all} connected components ${\cal C}'$ of $G_{\cal C}$ \textbf{do}\\
06 \hspace{2em} Let $C_{p_1}, \ldots, C_{p_k}$ be the vertices (cake subsets) in ${\cal C}'$; \\
07 \hspace{2em} \textbf{call} {\sc DC}$(\{ p_1, \ldots, p_k \}, C_{p_1} \cup \cdots \cup C_{p_k})$ \\
08 \hspace{2em} Let $C^*_{p_i}$ be the piece assigned by {\sc DC} in Line 07 for $i = 1, \ldots, k$; \\
09 \hspace{1em} \textbf{enddo} \\
10 \hspace{1em} \textbf{output} $C^*_p$ for every $p \in P_r$;\\
\textbf{end.}\\

\noindent
\textbf{procedure} {\sc Deposit}($p,P,C, \epsilon' , t$)\\
\textbf{begin}\\
01 \hspace{1em} $C' := C$, $h:= \left\lceil \frac{2^{10}t}{\epsilon'} \ln{\frac{1}{\epsilon'}} \right\rceil$\\
02 \hspace{1em} \textbf{from} $j=1$ \textbf{to} $54 \left( \ln{\frac{1}{\epsilon'}} \right)^2$  \textbf{do} \\
03 \hspace{2em} Choose a player from $P$ UAR $h$ times and let $P_0$ be the multiset of the chosen players;\\
04 \hspace{2em} \textbf{for all} $q \in P_0$ \textbf{do}\\
05 \hspace{3em} $\alpha_q := \mbox{\sc Eval}(C',q)$\\
06 \hspace{2em} \textbf{enddo} \\
07 \hspace{2em} Let $P'$ be the multiset of the players $q \in P_0$ such that $\alpha_q \geq \epsilon'$; \\
08 \hspace{2em} \textbf{if} $|P'| < 2^9 t \ln{\frac{1}{\epsilon'}}$ \textbf{then}\\
09 \hspace{3em} \textbf{output} $C'$; \textbf{return}\\
10 \hspace{2em} \textbf{endif} \\
11 \hspace{2em} \textbf{call} $\mbox{\sc Condense}(p,P',C')$\\
12 \hspace{1em} \textbf{enddo} \\
13 \hspace{1em} \textbf{return}\\
\textbf{end.}\\

\noindent
\textbf{procedure} {\sc Condense}($p,P',C'=[a,b]$)\\
\textbf{begin}\\
01 \hspace{1em} $x_L:=a$, $x_R:=b$\\
02 \hspace{1em} \textbf{for all} $q \in P'$ \textbf{do}\\
03 \hspace{2em} $\beta_q := \mbox{\sc Eval}([x_L,x_R],q)$ \\
04 \hspace{2em} $x_q := \mbox{\sc Cut}([x_L,x_R], q, \beta_p/2)$ \\
05 \hspace{1em} \textbf{enddo} \\
06 \hspace{1em} Let $q_0$ be the player such that $x_{q_0}$ is the median of multiset $\{ x_q ~|~ q \in P' \}$;\\ 
07 \hspace{1em} $\alpha_L := \mbox{\sc Eval}([x_L,x_{q_0}],p)$\\
08 \hspace{1em} $\alpha_R := \mbox{\sc Eval}([x_{q_0},x_R],p)$\\
09 \hspace{1em} \textbf{if} $\alpha_L > \alpha_R$ \textbf{then} \\
10 \hspace{2em} $C' := [x_L,x_{q_0}]$\\
11 \hspace{1em} \textbf{else}\\ 
12 \hspace{2em} $C' := [x_{q_0},x_R]$\\
13 \hspace{1em} \textbf{endif}\\ 
14 \hspace{1em} \textbf{return}\\
\textbf{end.}\\

$\widehat{C} := \cup_{p \in P_r} C_p$. 
The completion part of the algorithm for Theorem~\ref{th:S1} is simply applying {\sc Completion}($P-P_r,C-\widehat{C},n, \epsilon$).

\subsection{Proof of Theorem~\ref{th:S1}}

For $D \subseteq C$ and $0 \leq \alpha \leq 1$, we denote the set of players $p \in P$ such that $\mbox{\sc Eval}(D,p) \geq \alpha$ by $P(\alpha,D)$. 

\begin{lm}\label{lm:S1-1}
For $p \in P$, $D \subseteq C$,  and real numbers $0< \epsilon <1/e$ and $t \geq 1$, 
we choose players from $P(\epsilon,D)$ uniformly at random at least $2^9 t/\epsilon$ times and let $Q$ be the multiset of the chosen players. Let $D'$ denote the output $D$ of $\mbox{\sc Condense}(p,Q,D)$. Then the following two conditions hold:

\begin{itemize}
\item $\mbox{\sc Eval}(D',p) \geq \mbox{\sc Eval}(D,p)/2$, and 
\item $\mbox{\sc Eval}(D',q) \leq \mbox{\sc Eval}(D,q)/2$ for at least $|P(\epsilon,D)|/3$ players $q \in P(\epsilon,D)$ with probability at least $1 - \epsilon^{16 t}$.  
\end{itemize}
\end{lm}

\noindent
{\em Proof}: 
The first item ($\mbox{\sc Eval}(D',p) \geq \mbox{\sc Eval}(D,p)/2$) is clear from the operations in Lines 09-13 of {\sc Condense}. Then, we prove the second item. Let $|P(\epsilon,D)| = m$. Define $P_L = \mbox{\sc Pcut}(P(\epsilon,D), D, \mbox{\sc Eval}(D,q)/2, m/3)$ and $P_R = P - \mbox{\sc Pcut}(P(\epsilon,D), D, \mbox{\sc Eval}(D,q)/2, 2m/3)$. Let $Y_i^L$ (resp, $Y_i^R$) be a random variable such that it is 1 when the $i$th element of $Q$ is included in $P_L$ (rest., $P_R$) and 0 otherwise. $Y^L := \sum_{i=1}^{|Q|} Y_i^L$ and $Y^R := \sum_{i=1}^{|Q|} Y_i^R$. Clearly $E[Y^L] = E[Y^R] = |Q|/3 \geq {2^9 t}/{3 \epsilon}$. Every $Y_i^L$ and $Y_i^R$ is an independent Bernoulli trial, and thus from Chernoff bound ($\mbox{\rm Pr}[X \geq (1+ \delta) E[X]] \leq e^{-{\delta^2 E[X]}/{3}}$), it follows that 
$$ 
\mbox{\rm Pr}\left[Y^L \geq \frac{|Q|}{2} \right] = \mbox{\rm Pr}\left[Y^L \geq \frac{3}{2} E[Y^L] \right] \leq e^{-{E[Y^L]}/{12}}\leq e^{-{128 t}/{9 \epsilon}}
$$ 
Here, by considering that for all real number $x$, 
$$ 
x \ln{\frac{1}{x}} \leq \frac{1}{e} \leq \frac{4}{9}, 
$$
 we get 
$$ 
\mbox{\rm Pr}\left[Y^L \geq \frac{|Q|}{2} \right]  \leq e^{-32 t \ln{({1}/{\epsilon})}} = \epsilon^{32 t}.
$$

Similarly, we also get $\mbox{\rm Pr}\left[Y^R \geq {|Q|}/{2}\right] \leq \epsilon^{32 t}$. 
Let $q_0$ be the player in line 06 of {\sc Condense}$(p,Q,D)$. Then, $\mbox{\rm Pr}[q_0 \in P_L \cup P_R] \leq 2 \epsilon^{32 t} \leq \epsilon^{16 t}$. Therefore, for at least $m/3$ players $q$ (i.e., players in $P_L$), $\mbox{\sc Eval}([x_{q_0},x_R],q) \leq \mbox{\sc Eval}(D,q)/2$ and for at least $m/3$ players $q'$ (i.e., players in $P_R$), $\mbox{\sc Eval}([x_L,x_{q_0}],q') \leq \mbox{\sc Eval}(D,q')/2$ with probability at least $1- \epsilon^{16 t}$.  
\qed

\begin{lm}\label{lm:S1-2}
If  $P(\epsilon', C') \geq \epsilon' n$ when 
an operation of Line 08 of 
{\sc Deposit}$(p,P,C, \epsilon', t )$ is done, 
then the probability that 
$|P'| < 2^9 t \ln{\frac{1}{\epsilon'}}$ 
occurs 
is at most 
$\epsilon'^{128 t}$. 
\end{lm}

\noindent
{\em Proof}: 
Let $P_0 = \{ q_1, \ldots, q_h \}$ be $P_0$ constructed in Line 03 of {\sc Deposit}.
Let $X_i$ ($i = 1, \ldots, h$) be the random variable 
such that 
$X_i = 1$ if $q_i \in P(\epsilon', C')$ and 
$X_i = 0$ otherwise. 
Let $X$ be the random variable 
representing $|P_0 \cap P(\epsilon', C')|$. 
Clearly, $X = \sum_{i=1}^{h} X_i$ and 
\[
E[X] = \frac{|P(\epsilon', C')|}{n} \cdot \frac{2^{10}t}{\epsilon'} \ln \frac{1}{\epsilon'} \geq {2^{10}t} \ln \frac{1}{\epsilon'}. 
\]
From the Chernoff bound, 

\begin{eqnarray*}
{\mbox Pr}\left[ X \leq 2^9 t \ln \frac{1}{\epsilon'} \right] 
&=& {\mbox Pr} \left[ X \leq \left( 1- \frac12 \right) 2^{10} t \ln \frac{1}{\epsilon'} \right] \\
&\leq& {\mbox Pr} \left[ X \leq \left( 1- \frac12 \right) E[X] \right] \\
&\leq& e^{- {E[X]}/{8}}\\
&\leq& e^{2^7 t} \ln \frac{1}{\epsilon'}\\
&=& {\epsilon'}^{128 t}.
\end{eqnarray*}
\qed\\

In our algorithm, 
we call $\mbox{\sc Condense}(p,P',C')$ iteratively. Then, for distinguishing $C'$s in different calls, 
we number them such as 
$C^{(1)}$, $C^{(2)}$, $\ldots$: 
$C^{(1)}$ is $C'$ of the first call of $\mbox{\sc Condense}(p,P',C')$ 
(i.e., $C^{(1)} = C$), 
and the output of 
$\mbox{\sc Condense}(p,P',C^{(i)})$ is 
$C^{(i+1)}$ for $i \in \{ 1, 2, \ldots \}$. 
We say a call $\mbox{\sc Condense}(p,P',C^{(i)})$ is {\em good} if for at least $|P(\epsilon',C^{(i)})|/3$ players $q \in P(\epsilon',C^{(i)})$,
\begin{equation}
\mbox{\sc Eval}(C^{(i+1)},q) \leq  \mbox{\sc Eval}(C^{(i)},q)/2. \label{eq:S-1}
\end{equation}
From Lemma~\ref{lm:S1-1}, a call $\mbox{\sc Condense}(p,P',C^{(i)})$ is good with probability at least $1 - \epsilon'^{16 t}$.

\begin{lm}\label{lm:S1-3}
Assume that $C^{(j)}$ is obtained from $C^{(i)}$ after at least $\frac92 (\ln_{1/2}{\epsilon'} + 1)$ good calls. Then $|P(\epsilon',C^{(j)})| \leq \frac23 |P(\epsilon',C^{(i)})|$. 
\end{lm}

\noindent
{\em Proof}: 
Assume that $|P(\epsilon',C^{(j)})| > \frac23 |P(\epsilon',C^{(i)})|$. It is clear that $C^{(j)} \subseteq C^{(j-1)} \subseteq \cdots \subseteq C^{(i)}$. Let $m=|P(\epsilon',C^{(i)})|$.  Then, for every $C^{(k)}$ ($k \in \{ i, i+1, \ldots, j \}$), 
\begin{equation}
|P(\epsilon',C^{(k)})| > \frac23m. \label{eq:S-2}
\end{equation}
Here, assume that if (\ref{eq:S-1}) occurs for a player $q \in P(\epsilon',C^{(i)})$, then $q$ gets a ``stone.'' If a player gets $\log_{1/2}{\epsilon'} +1$ stones, then $\mbox{\sc Eval}(C',q) \leq \epsilon'$ and $q$ is removed from $P(\epsilon',C')$. If $\mbox{\sc Condense}(p,P',C^{(i)})$ is good, at least $|P(\epsilon', C^{(i)})|/3$ stones are distributed. By considering (\ref{eq:S-2}), after $\frac92 (\ln_{1/2}{\epsilon'} + 1)$ good calls, at least $\frac23m \cdot \frac13 \cdot \frac92 (\ln_{1/2}{\epsilon'} + 1) = m (\ln_{1/2}{\epsilon'} + 1)$ stones are distributed. Since one player can get $\ln_{1/2}{\epsilon'} + 1$ stones at most, every player gets $\ln_{1/2}{\epsilon'} + 1$ stones and has been removed from the $P(\epsilon',C^{(j)})$, contradiction. 
\qed

\begin{lm}\label{lm:S1-4}
{\em 
Let $C_p$ be the output of $\mbox{\sc Deposit}(p,P,C, \epsilon', t)$. Assume that $\epsilon' \leq 1/e$. Then 
\begin{itemize}
\item[(i)]  $\mu_p (C_p) \geq \left( \frac12 \right)^{54(\ln(1/\epsilon'))^2}$, and 
\item[(ii)] $|P(\epsilon',C_p)| \leq \epsilon' n$ with a probability of at least $1 - \epsilon'^{2t}$. 
\end{itemize}
}
\end{lm}

\noindent
{\em Proof}: 
From Lemma~\ref{lm:S1-1}, (i) is clear. 
Consider line 08 of $\mbox{\sc Deposit}(p,P,C, \epsilon', t)$. Assume that ${|P(\epsilon',C')|} \geq \epsilon' n$. Then, from Lemma~\ref{lm:S1-2}, $\mbox{\sc Condense}(p,P',C')$ is called in probability at least $1- {\epsilon'}^{128 t}$. From Lemma~\ref{lm:S1-1}, $\mbox{\sc Condense}(p,P',C')$ is good with probability at least $1 - \epsilon'^{16 t}$

From Lemma~\ref{lm:S1-3}, 
by the following number of good calls, we get  
$|P(\epsilon',C')| \leq \epsilon' n$. 
\begin{eqnarray*}
&&\frac92 (\ln_{1/2} \epsilon' + 1) (\ln_{2/3} \epsilon' + 1)
= \frac92 \left( \frac{\ln {\frac{1}{\epsilon'}}}{\ln 2} +1 \right) \left( \frac{\ln {\frac{1}{\epsilon'}}}{\ln {\frac32}} +1 \right)\\
&<& \frac92 \left( 2 \ln{\frac{1}{\epsilon'}} + 1 \right)  \left( 3 \ln{\frac{1}{\epsilon'}} + 1 \right) 
~\mbox{($\because$ $\ln 2>1/2$ and $\ln{3/2}>1/3$)}\\
&<& \frac92 \left( 3 \ln{\frac{1}{\epsilon'}} \right)  \left( 4 \ln{\frac{1}{\epsilon'}}  \right) 
~\mbox{($\because$ from $\epsilon' \leq 1/e$, $\ln \frac{1}{\epsilon'} \geq 1$)}\\
&=& 54 \left( \ln \frac{1}{\epsilon'} \right)^2. 
\end{eqnarray*}
The probability that ``$\mbox{\sc Condense}(p,P',C')$ is called and the call is good'' $54 \left( \ln \frac{1}{\epsilon'} \right)^2$ times in a row is at least 
\begin{eqnarray*}
&& 1-\left(\epsilon'^{128 t} + \epsilon'^{16 t} \right) \cdot 54 \left( \ln {1}/{\epsilon'} \right)^2\\
&\geq& 1-\left(\epsilon'^{8 t} \right) (1/\epsilon')^4 (1/\epsilon')^2 
~~\mbox{($\because$ $54 < e^4 \leq (1/\epsilon')^4$ and $\ln{1/\epsilon' } \leq 1/\epsilon'$)}\\
&\geq& 1 - \epsilon'^{2t}.
\end{eqnarray*}
Therefore (ii) is obtained. \qed\\

\noindent
{\em Proof of Theorem~\ref{th:S1}}: 
We will show the following facts: 
\begin{itemize}
\item[(i)] Each player in $P_r$ gets a fair portion
by  {\sc PreassignS}$(P,C,P_r,\epsilon,t)$. 
\item[(ii)]  $Q$ (the output of $\mbox{\sc Victimize}(P-P_r,C-\widehat{C},  \lfloor \epsilon n \rfloor )$ in line 01 of {\sc CompletionU}($P-P_r, C-\widehat{C}, n, \epsilon $) is safe with respect to $C-\widehat{C}$ with probability at least $1  - ({\epsilon/r})^{t} \geq 1  - {e}^{-t}$. 
\end{itemize}
In what follows, we show proofs of the above items. Note that it is sufficient to consider the case that
$\epsilon' = \epsilon/r$ in Lemmas~\ref{lm:S1-2}, \ref{lm:S1-3},  and \ref{lm:S1-4}. 
\begin{itemize}
\item[(i)] 
From Lemma~\ref{lm:S1-4}, $\mu_p(C_p) \geq (1/2)^{54 (\ln(r/\epsilon))^2}$ for every player $p \in P_r$. Thus by {\sc PreassignS}, every player finally gets a portion having at least $(1/2)^{54 (\ln(r/\epsilon))^2}/r$ value. 
We will show 
\begin{equation}
(1/2)^{54 (\ln(r/\epsilon))^2}/r \geq 1/n. \label{eq:S-3}
\end{equation}
This inequality can be transformed as follows: 
\begin{eqnarray}
r \cdot 2^{54 \left(\ln \left( \frac{r}{\epsilon} \right) \right)^2} &\leq& n  \nonumber\\
\ln r + 54 \ln 2 \cdot \left( \ln \left( \frac{r}{\epsilon} \right) \right)^2 &\leq& \ln n \nonumber
\end{eqnarray}
Here, from $\ln r \leq \ln (r/\epsilon) \leq (\ln (r/\epsilon ))^2$ ($\because$ $1/\epsilon \geq e$), the following inequalities hold: 
\begin{eqnarray*}
\ln r + 54 \ln 2 \cdot \left( \ln \frac{r}{\epsilon} \right)^2 \leq \left( 1 + 54 \ln 2 \right) \left( \ln \frac{r}{\epsilon} \right)^2 \leq \left( 7 \ln \frac{r}{\epsilon} \right)^2
\end{eqnarray*}
Thus, if $(7 \ln ({r}/{\epsilon}) )^2 \leq \ln n$, then (\ref{eq:S-3}) holds. This is equivalent to 
\begin{equation*} {r} \leq {\epsilon}  e^{\frac{\sqrt{\ln n}}{7}}.  
\end{equation*}
That is, (\ref{eq:S-3}) holds.  
\item[(ii)] 
For $p \in P_r$, 
if $|P(\epsilon/r, C_p)| \leq (\epsilon/r) n$, then we say that $p$ is {\em polite}. From Lemma~\ref{lm:S1-4}, the probability that $p  \in P_r$ in not polite is at most $(\epsilon/r)^{2t}$. Thus, the probability that at least one $p  \in P_r$ in not polite is at most $r (\epsilon/r)^{2t} \leq (\epsilon/r)^{t} \leq e^{-t}$ (since $\epsilon/r \leq \epsilon \leq 1/e$). 
If all players in $P_r$ are polite, then \[|P(\epsilon, \widehat{C})| \leq \sum_{p \in P_r} |P(\epsilon/r, C_p)| \leq r \cdot \frac{\epsilon}{r} n  = \epsilon n. \] Since $|P(\epsilon, \widehat{C})|$ is an integer, $|P(\epsilon, \widehat{C})| \leq \lfloor \epsilon n \rfloor$. Thus, all players in $P(\epsilon, \widehat{C})$ are removed by {\sc Victimize} with probability at least $1  - ({\epsilon/r})^{t} \geq 1  - {e}^{-t}$. 
\end{itemize}

It remains necessary to calculate the query complexity. In {\sc PreassignS}, {\sc Deposit} is called $r$ times and needs $$O \left( r \cdot \frac{rt}{\epsilon}  \ln \frac{r}{\epsilon} \cdot \left( \ln \frac{r}{\epsilon} \right)^2 \right)  = O\left( \frac{r^2 t}{\epsilon}  \left( \log \frac{r}{\epsilon} \right)^3 \right) $$ time. {\sc DC} for $k$ players can be done in $O(k \log k)$-time if a cake is continuous. However, {\sc DC}$(\{ p_1, \ldots, p_k \}, C_{p_1} \cup \cdots \cup C_{p_k})$ in line 07 of {\sc PreassignS} treats $C_{p_1} \cup \cdots \cup C_{p_k}$, which may be separated into at most $r$ continuous pieces. One query on a cake consisting of $k$ continuous pieces is simulated by $k$ queries on the continuous parts. Hence, the query complexity of this {\sc DC} is $O(r^2 \log (r^2))=O(r^2 \log r)$. Therefore, the time-complexity of {\sc PreassignS} is $O( ({r^2 t}/{\epsilon})  ( \log({r}/{\epsilon}) )^3 + r^2 \log r ) = O( ({r^2 t}/{\epsilon})  ( \log({r}/{\epsilon}) )^3)$.

For the completion part, {\sc DC}$(Q,C-\widehat{C})$ in {\sc Completion} is dominant. $C-\widehat{C}$ may be separated into at most $r+1$ continuous parts. Thus, the query complexity of {\sc DC} (and the completion part) is $O(rn \log (rn))$.  
\qed

\section{Summary}\label{sc:summary}

Herein, we considered a way to solve the cake-cutting problem in sublinear time. For this purpose, we introduced the concept of ``$\epsilon n$ victims,'' and presented the following framework. In the first (preassigning) part, we preassign operations to $r = o(n)$ players in $o(n)$ time. Then, in the second (completion) part, we assign portions to the remaining $n-r$ players except for the $\epsilon n$ victims in polynomial-time. (Note that the second part clearly requires $\Omega (n)$-time.) Within this framework, we presented two types of algorithms. In the first, the preassigned players cannot be designated, while in the second, they can be. 

For our future work, it remains necessary to show nontrivial lower-bounds. For example, we have not yet proven that only one victim is needed to preassign sublinear players in sublinear-time. Since numerous variations may be considered in our framework, the ability to make extended algorithms is also an attractive subject.

\section*{Acknowledgement}
We would like to thank Assistant Professor Yuichi Yoshida of the National Institute of Informatics for his valuable advice. We are also grateful for the ``Algorithms on Big Data'' project (ABD14) of CREST, JST, the ELC project (MEXT KAKENHI Grant Number 24106003), and JSPS KAKENHI Grant Numbers 24650006 and 15K11985, through which this work was partially supported. 



\begin{thebibliography}{99}
%
\bibitem{BSS_MC-testable_STOC08}
I. Benjamini, O. Schramm, and A. Shapira: 
Every minor-closed property of sparse
graphs is testable, Proc. STOC 2008, ACM, 2008, pp.~393--402. 
%
%
\bibitem{art11}
Brams, S. J. and Aran, D. T.: 
An envy-free cake division protocol, 
American Mathematical Monthly .
%
\bibitem{art5}
Edmonds, J. and Pruhs, K.: 
Cake cutting really isn't a piece of cake, 
Proceedings of the 17th Annual ACM-SIAM Symposium on Discrete Algorithms, 
Vol. 7 (2006).
%
\bibitem{art6}
Edmonds, J. and Pruhs, K.: 
Balanced Allocations of Cake, 
Proceedings of the 47th Annual IEEE Symposium on Foundations of Computer Science, 
pp. 623--634 (2006).
%
\bibitem{art2}
Even, S. and Paz, A.:  
A note on cake cutting, 
Discrete Applied Mathematics, 
Vol. 7, 
pp. 285--296 (1984). 
%
\bibitem{PropertyTestingLNCS10}
O. Goldreich (Ed.): 
Property Testing --- Current Research and Surveys, 
LNSC 6390, 2010. 
%
%
\bibitem{GR-STOC97}
O. Goldreich and D. Ron: 
Property testing in bounded
degree graphs: Proc. STOC 1997, 1997, pp.~406--415.
%
%
\bibitem{GGR-JACM98}
O. Goldreich, S.~Goldwasser, and D.~Ron: 
Property testing and its connection to learning and approximation: 
Journal of the ACM, Vol.~45, No.~4, July, 1998, pp.~653--750. 
%
%
\bibitem{HKNO_LocalPartition_FOCS09}
A.~Hassidim, J.~A.~Kelner, H.~N.~Nguyen, and K.~Onak:
Local graph partitions for approximation and testing, 
Proc. FOCS 2009, IEEE,  pp.~22--31.
%
%
\bibitem{ItoYoshida_Knapsack_TAMC12}
H.Ito, S. Kiyoshima, and Y.~Yoshida: 
Constant-time approximation algorithms for the knapsack problem, 
Proceedings of the 9th Annual Conference on TAMC, 
pp. 131--142 (2012).
%
\bibitem{ItoTanigawaYoshida_ICALP12}
H. Ito, S. Tanigawa, and Y. Yoshida:
Constant-Time Algorithms for Sparsity Matroids, 
Proc. ICALP (1), LNCS 7391, 2012, pp.~498--509.
%
\bibitem{NS_Testable_STOC11}
I. Newman and C. Sohler: 
Every property of hyperfinite graphs is testable, 
Proc. STOC 2011, ACM, 2011, pp.~675--784. 
%
%
\bibitem{Levi-Ron_PO_ICALP13}
R.~Levi and D.~Ron: 
A quasi-polynomial time partition oracle
for graphs with an excluded minor, 
Proc. ICALP 2013 (1), LNCS, 7965, Springer, 2013, 
pp.~709--720.
%
%
\bibitem{book1}
Robertson, J. and Webb, W.: 
Cake-Cutting Algorithms: Be Fair If You Can, 
A.~K.~Peters (1998).
%
%
\bibitem{art1}
Steinhaus H.: 
The Problem of fair division,
Econometrica, 
Vol. 16, 
pp. 101--104 (1948). 
%
\bibitem{Yoshida_STOC14}
Yuichi Yoshida:
A characterization of locally testable affine-invariant properties via decomposition theorems, 
Proc. STOC 2014, ACM, 2014, pp.~154--163. 
%
%
%
\end{thebibliography}


\end{document}